\documentclass[12pt,a4paper]{article}
\usepackage[T1]{fontenc}
\usepackage[utf8]{inputenc}
\usepackage{lmodern,microtype}
\usepackage[margin=27mm]{geometry}
\usepackage{amsmath,amssymb,amsthm,mathtools}
\usepackage{enumitem,booktabs,needspace}
\usepackage[authoryear,round]{natbib}

\usepackage{tikz,caption,placeins}
\usetikzlibrary{arrows.meta,calc,patterns}
\definecolor{figblue}{RGB}{37,80,112}
\definecolor{figwarm}{RGB}{144,104,58}
\tikzset{figaxis/.style={semithick,-{Stealth[length=4pt]}},
  payoff/.style={black,semithick},
  envelope/.style={figblue,thick,dash pattern=on 4pt off 2pt},
  guide/.style={black!45,thin,densely dotted},
  boundary/.style={black,semithick},
  figtext/.style={font=\small,align=center}}
\usepackage[hidelinks]{hyperref}
\usepackage{url}
\setlist[enumerate]{itemsep=3pt,topsep=5pt}
\newtheorem{proposition}{Proposition}
\newtheorem{lemma}{Lemma}[section]
\newcommand{\E}{\mathbb{E}}
\newcommand{\Prb}{\mathbb{P}}
\newcommand{\Var}{\operatorname{Var}}
\newcommand{\cav}{\operatorname{cav}}
\newcommand{\ind}{\mathbf{1}}
\newcommand{\dd}{\,\mathrm{d}}

\allowdisplaybreaks[1]
\hypersetup{pdftitle={Persuasion under the Threat of Verification},pdfauthor={Georgy Lukyanov and Samuel Safaryan},pdfkeywords={Bayesian persuasion, verification, information acquisition, information sharing, receiver heterogeneity}}
\title{Persuasion under the Threat of Verification\thanks{We thank Olivier Gossner, Johannes H\"orner, Yukio Koriyama, Yves Le Yaouanq, Alessandro Riboni, Fran\c{c}ois Salani\'e, Anna Sanktjohanser, Ludvig Sinander, Nikita Zakharov, participants at the PEDD 2023 Conference and the TSE Applied Theory Workshop, and seminar participants at CREST for useful comments and discussions. Mira Hajar provided excellent research assistance. Any errors are our own.}}
\author{Georgy Lukyanov\thanks{Toulouse School of Economics, Toulouse, France. Corresponding author: \href{mailto:georgy.lukyanov@tse-fr.eu}{georgy.lukyanov@tse-fr.eu}.}
\and Samuel Safaryan\thanks{HSE University, International College of Economics and Finance, Moscow, Russia. \href{mailto:sesafaryan@edu.hse.ru}{sesafaryan@edu.hse.ru}.}}
\date{11 October 2026}
\begin{document}
\maketitle
\begin{abstract}
This paper studies persuasion when public information is costly and receivers can verify the state before acting. Verification may overturn acceptance after favorable public news, but it may also induce acceptance after unfavorable news. We characterize how these two effects shape the sender's optimal experiment. Cheaper verification can lead the sender to disclose less and rely more on receivers' own discoveries, thereby increasing its payoff. With heterogeneous decision thresholds, we identify exact conditions for the disclosure response and show why it depends on the distribution of preferences. In a separate parameter region, cheaper verification increases inspection but reduces every threshold type's final information and welfare; the sender also loses. We then allow receivers to share their findings. Cheaper inspection and wider sharing can move optimal public disclosure in opposite directions, even when both benefit the sender. Wider sharing can also reduce aggregate final information while improving every threshold type's welfare. The results distinguish the effects of verification on public disclosure, information ultimately obtained, and the cost of obtaining it.

\end{abstract}

\clearpage

\section{Introduction}\label{sec:intro}
Individuals often rely on information provided by a party with an interest in their decisions. A firm discloses evidence about a product that it wishes consumers to buy; an organization releases an assessment to an audience whose support it seeks. The audience may, however, check this information before acting. A public message then affects both the immediate decision and the incentive to acquire further evidence. How does this possibility change the sender's willingness to provide information, and do receivers necessarily benefit when verification becomes cheaper?

The effect of verification on the sender is not unambiguous. A receiver who would accept on the basis of public news may inspect, discover a bad state, and reject. A receiver who would otherwise reject may instead discover a good state and accept. In the first case, verification works against the sender's recommendation. In the second, it generates acceptance that the sender need not secure through public disclosure. When public information is costly, the sender may therefore wish to leave part of the learning to the audience. Whether this is profitable depends on the news that receivers initially receive and on their willingness to accept.

We study these questions in a binary-state persuasion model. The sender values acceptance and pays a cost proportional to the variance of public posterior beliefs. Receivers differ in their privately known costs of perfectly verifying the state. We begin with a common acceptance threshold and then allow thresholds to differ across receivers. The sender commits to a public experiment, receivers observe its realization and decide whether to inspect, and they finally choose their actions. In an extension, an inspection result reaches the whole audience with a given probability. Receivers may then rely on one another's findings as well as on the sender's information.

With a common threshold, we characterize the optimal experiment explicitly. When public information is sufficiently inexpensive, the sender combines a conclusive rejection signal with a posterior at which receivers accept. Cheaper verification can raise the acceptance posterior, making the public experiment more informative. When public information is more expensive, the favorable posterior stays at the acceptance threshold, while the unfavorable posterior may be positive. In this case, cheaper verification can raise the unfavorable posterior and reduce public information. Receivers who hear unfavorable news have a greater chance of discovering the good state, and the sender spends less on disclosure. Its payoff can rise even though the additional value of public persuasion, relative to providing no information, falls.

We next examine how this comparison changes when receivers have different thresholds. For uniformly distributed thresholds, the gains and losses in acceptance caused by verification cancel exactly. A smooth family of threshold distributions allows us to move away from this case and characterize the optimal experiment globally. Within this family, cheaper verification can expand, contract, and then again expand public information as its cost relative to threshold concentration rises. The common-threshold benchmark therefore does not provide a general cost boundary for the direction of the disclosure response. We also obtain a nonempty region in which cheaper verification reduces public information and benefits the sender.

The welfare implications require a separate comparison. More frequent inspection need not compensate receivers for a less informative public experiment. For every prior below the common acceptance threshold, we identify an open range of costs for which cheaper verification eliminates public disclosure and makes every threshold type worse informed in the Blackwell order. This result survives arbitrarily small atomless dispersion of thresholds and holds for every initial optimal experiment. The information and welfare comparisons average over private inspection costs within each threshold type. Inspection becomes more frequent, but expenditure falls and decisions deteriorate enough to reduce welfare. The sender also loses in this region. Thus the receivers' loss can arise even when the sender does not gain from the change in verification costs.

Sharing adds another consideration. A receiver who expects to learn from someone else's finding has less reason to inspect. We establish uniqueness of the continuation equilibrium and then solve the sender's problem for sufficiently small sharing probabilities. In a region where cheaper inspection reduces public disclosure, wider sharing increases it; both changes nevertheless benefit the sender. We also show that wider sharing may reduce the aggregate informativeness of final beliefs while improving every threshold type's welfare. For this last comparison, we give a sufficient condition under which providing no public information is optimal throughout. The comparison therefore allows the sender to choose its experiment. It concerns aggregate information, with preference labels suppressed, and does not assert that every threshold type becomes worse informed.

Taken together, these results show why the frequency of checks is insufficient for evaluating a verification technology. One also needs to account for the sender's disclosure response, the information obtained by receivers with different preferences, and their acquisition expenditure. The typewise losses from cheaper verification and the welfare gains from sharing occur in different regions of the model. They identify distinct effects that an assessment of either intervention must allow for.

The rest of the paper is organized as follows. Section~\ref{sec:model} presents the model and the information comparisons. Section~\ref{sec:benchmark} solves the common-threshold benchmark. Section~\ref{sec:heterogeneity} studies heterogeneous thresholds and the welfare effects of cheaper verification. Section~\ref{sec:sharing} introduces sharing, and Section~\ref{sec:discussion} discusses the scope of the results. Proofs and further analytical results are in the appendix.

\subsection{Related literature}\label{sec:literature}
Our paper builds on the Bayesian persuasion framework of \citet{KG2011} and its extension to costly information provision by \citet{GK2014}. The sender chooses a Bayes-plausible distribution of posterior beliefs. With our quadratic information cost, its problem can be solved by concavifying the continuation payoff after subtracting a quadratic term. We use this standard approach to obtain explicit optimal experiments and to study how they respond to the receivers' verification incentives. \citet{CS2025} develop more general comparative statics of persuasion using properties of payoff differences. Here the continuation payoff permits a direct characterization of the regions in which the optimal experiment expands or contracts.\footnote{The binary-prior comparison in \citet[Proposition~1]{CS2025} uses a weak order between sets when the optimum is not unique. It should not be read as a strict ordering of every pair of optimal experiments.}

The closest paper is \citet{MM2023}, who allow a receiver to acquire additional information after observing the sender's signal. The sender can provide information without paying a direct information cost, and the receiver has a flexible acquisition technology satisfying their assumptions. They show that there exists an optimal persuasion policy under which the receiver undertakes no further learning. The sender can consolidate the information that the receiver would otherwise acquire into its own experiment. In our setting, such consolidation has a direct cost to the sender. Receivers also face a restricted choice between no inspection and perfect verification, and a common public experiment must accommodate their different thresholds and private costs. These differences allow verification to occur in equilibrium. Indeed, under our verification technology it can occur even when the public information cost is zero. Their Proposition~3 further shows that the sender's equilibrium payoff weakly increases with the receiver's acquisition cost. Our sender can instead benefit from a reduction in verification costs, because favorable private discoveries can replace some costly public disclosure.

The comparison with \citet{MM2023} also helps clarify the welfare result. In their binary-state, binary-action case, receiver welfare weakly improves with cheaper acquisition under the assumptions of their Proposition~4; with richer action spaces, adverse effects are possible. We obtain a loss of final information and welfare for every threshold type with binary states and binary actions. The result depends on the different acquisition technology and the sender's public information cost: cheaper verification leads the sender to withdraw public information, and the additional private checks do not replace what receivers lose. Our smooth family of threshold distributions further shows how the disclosure response varies with the distribution of preferences.

\citet{Yang2026} studies endogenous state verification in a continuous-state persuasion model. Verification occurs on the equilibrium path, and the sender designs information to discourage checks that would overturn acceptance. In that model the sender weakly benefits when verification becomes more costly or less likely to succeed. With a uniform state distribution, cheaper or more effective verification weakly improves the probability of a correct final action. This is a decision-accuracy comparison, rather than a Blackwell ordering or a comparison that includes verification expenditure. Our model incorporates a direct cost of public disclosure and heterogeneous receivers. It allows us to characterize when discoveries following unfavorable public news are sufficiently valuable that the sender benefits from cheaper verification. We then trace the resulting changes in disclosure through to final information and welfare. The distinction concerns these incentives and their consequences, rather than the presence of endogenous verification itself.\footnote{Other verification technologies address different problems. In \citet{Zhou2023}, verification reveals the information plan actually implemented by a partially committed sender. \citet{RS2025} experimentally study a device that probabilistically flags false messages. Here the sender commits to its public experiment, while receivers choose whether to pay for evidence about the state.}

The timing of acquisition is also important. In \citet{PRV2026}, the receiver chooses a verification probability after observing the sender's information policy but before observing its realization. In our model, receivers see the realized public posterior before deciding whether to inspect. Their acquisition decisions can therefore respond to the content of the public news. \citet{BRV2020} study testing, disclosure, and approval, and show that incentives may favor an inferior testing technology. We hold the content of verification fixed and vary its cost or the dissemination of its findings, allowing the sender to adjust the public experiment in each case.

Heterogeneity also matters in \citet{BCG2024}, where a firm publicly discloses product information to consumers with different prior beliefs who can subsequently acquire further information at a cost. Their analysis shows how market heterogeneity and access to outside information affect disclosure. Our receivers share a prior but differ in decision thresholds and private inspection costs. We characterize how the threshold distribution affects the response to cheaper verification and compare final information and welfare within each threshold type.

Several papers examine how information provision affects the demand for further learning. \citet{Cheng2023} studies how freely provided information can deter, divert, or encourage costly learning. In the targeted-advertising model of \citet{MS2023}, information provision can encourage pessimistic consumers to learn and discourage optimistic consumers from doing so. Our favorable-discovery effect is related to this mechanism. We study it under a common public experiment and obtain exact conditions for its effect on disclosure and the sender's payoff. \citet{Hang2026} considers information and monetary incentives for private learning, whereas our sender uses only the public experiment. \citet{Wei2021} and \citet{DA2026} study costly learning or attention directed toward persuasive information. In \citet{Liao2021}, the sender chooses a single experiment from which the receiver can acquire costly independent signals over time. The costly information in our model is independent, fully revealing evidence obtained after the public signal. \citet{Ui2022} studies public disclosure with endogenous private acquisition in a different strategic environment. Our analysis adds explicit comparisons of the final information obtained by different preference types and of their acquisition costs.

The sharing extension relates to persuasion with several information-acquiring receivers. \citet{Montes2025} establishes equilibrium existence in persuasion games with multiple receivers and costly acquisition. For the private-cost verification game considered here, we obtain a unique continuation equilibrium, which allows us to solve the sender's disclosure problem without an equilibrium-selection assumption. \citet{Levkun2025} studies communication with strategic fact-checking, including free riding among fact-checkers. \citet{GTX2026} examine political persuasion with heterogeneous voters and an imperfect fact-checker. In our model the receivers themselves decide whether to inspect for the purpose of choosing their own actions, and their findings may subsequently reach other receivers. This specification makes it possible to compare a reduction in inspection costs with an increase in sharing.

More broadly, the public-good aspect of information acquisition is central to \citet{BK2007} and \citet{GG2010}. Access to others' information can crowd out private acquisition and concentrate information production. \citet{AJ2024} study acquisition and diffusion in a market setting, while \citet{MMS2025} examine proxy-advice design with information acquisition and free riding. We combine this acquisition problem with the choice of an upstream sender who pays for public information. The sender can respond differently to cheaper inspection and wider sharing, and the aggregate information effect of sharing can differ from its welfare effect. The appendix also studies how the sender's disclosure changes as the audience grows when each inspection can inform everyone.

\section{Model and information comparisons}\label{sec:model}
We consider one sender and $n\geq2$ receivers. The state is $\theta\in\{0,1\}$ with prior $\Prb(\theta=1)=\pi\in(0,1/2)$. Receiver $i$ chooses $a_i\in\{0,1\}$. The sender's gross payoff is the average action, $n^{-1}\sum_i a_i$. Each receiver privately observes a threshold $t_i\in(0,1)$ and an inspection cost $c_i$. The pairs $(t_i,c_i)$ are independent across receivers and independent of the state. The threshold distribution is $F$, and costs are independent of thresholds with
\begin{equation}\label{eq:costs}
c_i\sim U[0,1/r],\qquad 0<r\leq2.
\end{equation}
A larger $r$ makes verification cheaper in the sense of first-order stochastic dominance. We use $r=0$ for the limit in which no receiver verifies. The sender observes neither the thresholds nor the costs.

Before learning the state and before any receiver acquires information, the sender commits to a public experiment. We represent it by its induced distribution $\Pi$ of posterior beliefs $\mu=\Prb(\theta=1\mid\text{public signal})$. Feasibility requires
\begin{equation}\label{eq:bayes}
\int_0^1\mu\dd\Pi(\mu)=\pi.
\end{equation}
The sender pays $\kappa\Var_\Pi(\mu)$, where $\kappa\geq0$. This is a cost of the sender's experiment; receivers pay separately for any subsequent inspection.\footnote{For a binary state, $\Var_\Pi(\mu)=\pi(1-\pi)-\E_\Pi[\mu(1-\mu)]$. The cost is therefore proportional to the expected reduction in posterior variance. It is zero for an uninformative experiment.}

After observing the public posterior, receivers simultaneously decide whether to inspect. Inspection reveals $\theta$ perfectly and costs $c_i$. Initially, its outcome remains private. Section~\ref{sec:sharing} adds automatic sharing after these simultaneous decisions and before actions are taken. Writing $z_i$ for the inspection indicator, receiver $i$ has utility
\begin{equation}\label{eq:utility}
u_i=-2t_i a_i(1-\theta)-2(1-t_i)(1-a_i)\theta-c_i z_i.
\end{equation}
Without further information, the receiver accepts if $\mu\geq t_i$. At indifference we select acceptance. This convention matters for the common-threshold benchmark; it is immaterial under an atomless $F$.\footnote{The factor two makes the loss from an incorrect action equal to one when $t_i=1/2$. It does not change the economic content of threshold heterogeneity.}

\subsection{Independent verification}\label{sec:independent}
The value of learning the state at public posterior $\mu$ is
\begin{equation}\label{eq:q}
q(t,\mu)=2\min\{(1-t)\mu,t(1-\mu)\}.
\end{equation}
It never exceeds $1/2$, so \eqref{eq:costs} implies that the receiver inspects precisely when $c\leq q(t,\mu)$, with probability $rq(t,\mu)$. Conditional on $t$ and $\mu$, expected acceptance is
\begin{equation}\label{eq:action}
\ind\{t\leq\mu\}+rq(t,\mu)\bigl[\mu-\ind\{t\leq\mu\}\bigr].
\end{equation}
The second term is positive for a receiver who would reject the public recommendation and negative for a receiver who would accept it.

Let $s(\mu)$ denote aggregate expected acceptance. Under an atomless threshold distribution,
\begin{equation}\label{eq:s-general}
s(\mu)=F(\mu)+rH(\mu),\qquad
H(\mu)=\int\bigl[\mu-\ind\{t\leq\mu\}\bigr]q(t,\mu)\dd F(t).
\end{equation}
The sender chooses $\Pi$ to maximize
\begin{equation}\label{eq:sender}
V=\sup_{\Pi:\E\mu=\pi}\left\{\E_\Pi s(\mu)-\kappa\Var_\Pi(\mu)\right\}
=\cav v(\pi)+\kappa\pi^2,\qquad v(\mu)=s(\mu)-\kappa\mu^2.
\end{equation}
Here $\cav v$ is the smallest concave majorant of $v$. Appendix~\ref{app:prelim} explains how to implement the posterior law and solve the problem using supporting lines. Under independent verification, the expected average action is independent of $n$. The sender therefore has the same objective as with a continuum of independent receivers; no law-of-large-numbers assumption is needed.

For later use define
\begin{align}\label{eq:QAH}
A(\mu)&=\int_0^\mu t\dd F(t),& B(\mu)&=\int_\mu^1(1-t)\dd F(t),\\
Q(\mu)&=\int q(t,\mu)\dd F(t)=2[(1-\mu)A(\mu)+\mu B(\mu)],&
H(\mu)&=2[\mu^2B(\mu)-(1-\mu)^2A(\mu)].\nonumber
\end{align}
If $F$ has density $f$, then $Q'=2(B-A)$ and $Q''=-2f$. Setting $z(\mu)=\mu(1-\mu)$ gives the useful identity
\begin{equation}\label{eq:Hidentity}
H=zQ'-(1-2\mu)Q=z^2(Q/z)'.
\end{equation}
It relates the aggregate action effect of verification to the distribution of the value of learning.

\subsection{Public information, final information, and welfare}\label{sec:comparisons}
For $a\leq\pi\leq b$, let $\Pi[a,b]$ place probabilities $(b-\pi)/(b-a)$ and $(\pi-a)/(b-a)$ on $a$ and $b$. If an endpoint equals $\pi$, this denotes the uninformative law $\delta_\pi$. Its variance is $(\pi-a)(b-\pi)$. For two such experiments, inclusion of the smaller support interval in the larger one gives a Blackwell ordering: expanding the interval makes public information more informative.

A receiver's final posterior may differ from the public posterior. Averaging over inspection costs within threshold type $t$, its conditional distribution is
\begin{equation}\label{eq:finalkernel}
\mathcal K_t(\mu)=[1-rq(t,\mu)]\delta_\mu
+rq(t,\mu)[(1-\mu)\delta_0+\mu\delta_1].
\end{equation}
We write $\Xi_t=\int\mathcal K_t(\mu)\dd\Pi(\mu)$ for the final posterior law of type $t$. These laws all have mean $\pi$. A law $\Xi'$ is Blackwell more informative than $\Xi$ if and only if
\begin{equation}\label{eq:call}
C_{\Xi'}(x):=\int(\mu-x)_+\dd\Xi'(\mu)
\ \geq\ C_\Xi(x)\quad\text{for every }x\in[0,1].
\end{equation}
Strictness means dominance without equality of the two laws. We use this criterion to compare what receivers ultimately learn after both public disclosure and private verification.

We also use the aggregate final law $\Xi=\int\Xi_t\dd F(t)$. It describes the final posterior of a randomly selected receiver with the threshold label suppressed. An aggregate Blackwell comparison need not hold separately for each threshold type. Nor does it by itself order welfare when the preference distribution and the allocation of information across preferences matter.\footnote{The receiver knows its own threshold and cost. The cost-averaged law for a fixed threshold is a valid posterior representation because these types are independent of the state. The comparison concerns the ex ante experiment for that threshold, before the cost is drawn; it is not a claim about every realized cost.}

At $t$ and $\mu$, the probability of inspection, expected inspection expenditure, and total expected loss are respectively
\begin{equation}\label{eq:ind-welfare}
\Lambda_t=rq,\qquad E_t=\frac r2q^2,\qquad
L_t=q-\frac r2q^2.
\end{equation}
The decision loss alone is $q-rq^2$. Ex ante quantities integrate these expressions over $\Pi$; welfare is the negative of total loss. A welfare comparison must therefore account for both mistakes and inspection expenditure, including when the final experiments can be Blackwell ordered.

\section{A common decision threshold}\label{sec:benchmark}
Suppose every receiver has threshold $t=1/2$. Expected acceptance is
\begin{equation}\label{eq:benchmark-s}
s(\mu)=\begin{cases}
r\mu^2,&0\leq\mu<1/2,\\
1-r(1-\mu)^2,&1/2\leq\mu\leq1.
\end{cases}
\end{equation}
Below the threshold, only receivers who inspect and discover the good state accept. Above it, verification overturns acceptance only in the bad state. The upward jump at $1/2$ has size $1-r/2$. Cheaper verification both reduces this jump and raises the payoff from unfavorable public posteriors.

The optimal experiment can take two forms. When public information is inexpensive, the sender combines a conclusive rejection signal with an acceptance posterior at or above $1/2$. When information is more expensive, the sender can reduce the dispersion of posteriors by keeping the favorable posterior at $1/2$ and moving the unfavorable posterior above zero. Receivers may then discover the good state and accept even after unfavorable public news. Figure~\ref{fig:envelopes} illustrates the two experiments. The following proposition characterizes the regions in which they are optimal and their responses to cheaper verification.

\begin{figure}[t]
\centering
% Schematic drawing coordinates are not economic parameter values.
\begin{minipage}[t]{.49\linewidth}
\centering
\textbf{(a)}\quad$\Pi[0,y]$, $y>1/2$\par\smallskip
\begin{tikzpicture}[x=.92cm,y=.92cm,font=\small]
\path[use as bounding box] (-.45,-.25) rectangle (6.6,4.55);
\draw[figaxis] (0,.35)--(6.35,.35) node[right] {$\mu$};
\draw[figaxis] (0,.1)--(0,4.15) node[above] {$v,\ \cav v$};
\draw[payoff,domain=0:3,samples=41] plot (\x,{.35+(.65+2*.27*.7)/6*\x*\x});
\draw[payoff,domain=3:6,samples=61] plot (\x,{.35+.65*\x-.27*(\x-3.7)^2});
\draw[envelope] (0,.35)--(3.7,2.755);
\draw[envelope,domain=3.7:6,samples=41] plot (\x,{.35+.65*\x-.27*(\x-3.7)^2});
\draw[guide] (1.45,.35)--(1.45,1.2925);
\draw[guide] (3,.35)--(3,2.1677);
\draw[guide] (3.7,.35)--(3.7,2.755);
\filldraw[fill=white,draw=black] (3,1.892) circle (2pt);
\fill (3,2.1677) circle (1.8pt);
\fill[figblue] (0,.35) circle (2.3pt);
\fill[figblue] (3.7,2.755) circle (2.3pt);
\foreach \x/\lab in {0/0,1.45/\pi,3/\frac12,3.7/y,6/1}
 {\draw (\x,.35)--(\x,.27);\node[below] at (\x,.24) {$\lab$};}
\node[figtext,text=figblue] at (2.0,3.55) {supporting chord};
\draw[figblue,thin,-{Stealth[length=3pt]}] (2.0,3.27)--(2.35,1.8775);
\end{tikzpicture}
\end{minipage}\hfill
\begin{minipage}[t]{.49\linewidth}
\centering
\textbf{(b)}\quad$\Pi[\ell,1/2]$, $0<\ell<\pi$\par\smallskip
\begin{tikzpicture}[x=.92cm,y=.92cm,font=\small]
\path[use as bounding box] (-.45,-.25) rectangle (6.6,4.55);
\draw[figaxis] (0,3.75)--(6.35,3.75) node[right] {$\mu$};
\draw[figaxis] (0,.1)--(0,4.15) node[above] {$v,\ \cav v$};
\draw[payoff,domain=0:3,samples=41] plot (\x,{3.75-.09*\x*\x});
\draw[payoff,domain=3:6,samples=41] plot (\x,{3.75-.5184-.54*(\x-3)-.14*(\x-3)^2});
\draw[envelope,domain=0:1.2,samples=21] plot (\x,{3.75-.09*\x*\x});
\draw[envelope] (1.2,3.6204)--(3,3.2316);
\draw[envelope,domain=3:6,samples=41] plot (\x,{3.75-.5184-.54*(\x-3)-.14*(\x-3)^2});
\draw[guide] (1.2,3.75)--(1.2,3.6204);
\draw[guide] (2.05,3.75)--(2.05,3.4368);
\draw[guide] (3,3.75)--(3,2.94);
\filldraw[fill=white,draw=black] (3,2.94) circle (2pt);
\fill[figblue] (1.2,3.6204) circle (2.3pt);
\fill[figblue] (3,3.2316) circle (2.3pt);
\node[left] at (-.02,3.75) {$0$};
\foreach \x/\lab in {1.2/\ell,2.05/\pi,3/\frac12,6/1}
 {\draw (\x,3.75)--(\x,3.83);\node[above] at (\x,3.86) {$\lab$};}
\node[figtext,text=figblue] at (1.55,1.30) {supporting chord};
\draw[figblue,thin,-{Stealth[length=3pt]}] (1.7,1.58)--(2.35,3.372);
\end{tikzpicture}
\end{minipage}
\caption{The two forms of the informative benchmark experiment. Solid curves show the adjusted payoff $v$, and dashed curves its concave envelope. Filled blue points give the optimal public posteriors, with $\pi$ between them. Panel (a) applies when $\kappa<2$ and $(\kappa+4)/3<r<2$; panel (b) when $\kappa>2$ and $0<\ell<\pi$. The open point is the excluded left limit at $1/2$. As verification becomes cheaper, $y$ moves right in panel (a), expanding public information, while $\ell$ moves right in panel (b), contracting it.}
\label{fig:envelopes}
\end{figure}
\FloatBarrier

\begin{proposition}\label{prop:benchmark}
Let all thresholds equal $1/2$ and put $D=1/2-\pi$.
\begin{enumerate}[label=(\roman*)]
\item If $0\leq\kappa<2$, the unique optimal experiment is $\Pi[0,y]$, where
\begin{equation}\label{eq:ybench}
y=\begin{cases}1/2,&r\leq(\kappa+4)/3,\\
\sqrt{(r-1)/(r+\kappa)},&r>(\kappa+4)/3.
\end{cases}
\end{equation}
It becomes weakly more informative as $r$ rises, and strictly so in the second region. Sender value strictly decreases with $r$.
\item If $\kappa>2$, the unique optimal experiment is $\Pi[\ell,1/2]$, where
\begin{equation}\label{eq:ellbench}
\ell=\min\left\{\pi,\max\left\{0,\frac12-\sqrt{\frac{1-r/2}{\kappa-r}}\right\}\right\}.
\end{equation}
It becomes weakly less informative as $r$ rises, and strictly so when $0<\ell<\pi$. In that interior region sender value rises with $r$ exactly when
\begin{equation}\label{eq:ellcrit}
\ell>\ell_c:=\frac{\sqrt{1+16D\pi}-1}{8D}.
\end{equation}
\item If $\kappa=2$ and $r<2$, the unique optimum is $\Pi[0,1/2]$. At $\kappa=r=2$, every mean-$\pi$ law supported on $[0,1/2]$ is optimal.
\end{enumerate}
\end{proposition}

In the first region, raising $y$ makes a favorable recommendation less vulnerable to a bad-state discovery. The sender pays for this change in two ways: favorable recommendations become less frequent, and the public experiment becomes more expensive. In the second region, raising $\ell$ makes unfavorable public news less conclusive. This raises the chance that receivers inspect and discover the good state, while reducing the public information cost. Favorable recommendations again become less frequent, so the sender does not gain throughout the contraction region.

The trade-off is explicit if gross acceptance is decomposed into baseline acceptance, favorable discoveries, and overturned acceptances:
\begin{equation}\label{eq:decomp}
V=\underbrace{\Prb(\mu\geq1/2)}_{B}
+\underbrace{r\E[\mu^2\ind\{\mu<1/2\}]}_{R}
-\underbrace{r\E[(1-\mu)^2\ind\{\mu\geq1/2\}]}_{O}
-\kappa\Var(\mu).
\end{equation}
For $\Pi[0,y]$, there are no favorable discoveries after the unfavorable signal: $R=0$. The derivative of value with respect to its acceptance posterior, holding $r$ fixed, is
\begin{equation}\label{eq:dy-economic}
\frac{\partial V}{\partial y}=\pi\left[\frac{r-1}{y^2}-(r+\kappa)\right].
\end{equation}
For $\Pi[\ell,1/2]$, writing $h_0=1/2-\ell$, the corresponding derivative is
\begin{equation}\label{eq:dell-economic}
\frac{\partial V}{\partial\ell}
=\frac{D}{h_0^2}\left[-1+\frac r2+(\kappa-r)h_0^2\right].
\end{equation}
At an interior optimum, these marginal effects balance. As the lower posterior rises, the two public posteriors become less dispersed. Once $\ell=\pi$, the sender provides no public information.

To see why the sender can gain while providing less information, consider its payoff within the interior contraction region:
\begin{equation}\label{eq:premium}
V=r\pi^2+
\left[\sqrt{1-r/2}-D\sqrt{\kappa-r}\right]^2.
\end{equation}
The first term is the payoff the sender would receive without providing information. The second is the additional payoff from its optimal public experiment. The latter strictly decreases with $r$, even in the region where total sender value increases. The sender gains because cheaper verification generates more acceptance in the absence of public disclosure, while the advantage of providing public information becomes smaller.

Increasing $\kappa$ has a different interpretation. For $r<2$, optimal public information contracts weakly as its cost rises. In the active expansion region, $y_\kappa=-y/[2(r+\kappa)]<0$; in the active contraction region, $\ell_\kappa=\sqrt{1-r/2}/[2(\kappa-r)^{3/2}]>0$. Verification frequency weakly increases along these changes, but total receiver loss also weakly increases. The additional inspection does not compensate receivers for the reduction in public information. The envelope formula is $V_\kappa=-\Var_\Pi(\mu)$ wherever the derivative exists. It orders sender value but does not imply that actual expenditure $\kappa\Var_\Pi(\mu)$ is increasing in $\kappa$.

The no-verification limit illustrates the role of public information costs. At $r=0$, the optimum is $\Pi[0,1/2]$ for $\kappa\leq4$, is $\Pi[1/2-1/\sqrt\kappa,1/2]$ for $4<\kappa<D^{-2}$, and is pooling for $\kappa\geq D^{-2}$. The corresponding values are $2\pi-\kappa\pi D$, $(1-D\sqrt\kappa)^2$, and zero. Receiver loss equals $\pi$ throughout. At the other limit, $\kappa=0$, verification still occurs on the equilibrium path when $r>0$. Thus positive public information costs are not necessary for verification to occur under the restricted technology considered here.

Figure~\ref{fig:benchmark-regions} collects the benchmark support regions. Its three boundaries come from the upper tangency condition, the lower contact reaching zero, and the lower contact reaching the prior. The location of the pooling boundary depends on the prior even though the reversal between the two active responses occurs at $\kappa=2$.

\begin{figure}[t]
\centering
% Symbolic intercepts; horizontal distances involving D^{-2} are schematic.
\begin{tikzpicture}[x=1.35cm,y=2.80cm,font=\small]
\fill[black!4] (0,0)--(4,0)--(2,2)--(0,{4/3})--cycle;
\fill[figblue!12] (0,{4/3})--(0,2)--(2,2)--cycle;
\fill[figwarm!13] (2,2)--(4,0)--(7.3,0)--cycle;
\fill[black!10] (2,2)--(8.6,2)--(8.6,0)--(7.3,0)--cycle;
\draw[guide] (0,2)--(8.6,2);
\draw[guide] (2,0)--(2,2);
\draw[boundary] (0,{4/3})--(2,2);
\draw[boundary] (2,2)--(4,0);
\draw[boundary] (2,2)--(7.3,0);
\draw[figaxis] (0,0)--(8.95,0) node[right] {$\kappa$};
\draw[figaxis] (0,0)--(0,2.2) node[above] {$r$};
\foreach \x/\lab in {0/0,2/2,4/4,7.3/D^{-2}}
 {\draw (\x,0)--(\x,-.04);\node[below] at (\x,-.05) {$\lab$};}
\foreach \y/\lab in {1.333333/{4/3},2/2}
 {\draw (0,\y)--(-.07,\y);\node[left] at (-.08,\y) {$\lab$};}
\node[figtext] at (1.25,.60) {$\Pi[0,1/2]$\\unchanged support};
\node[figtext,text=figblue] at (.66,1.79) {{$\Pi[0,y]$}\\ {$y_r>0$}};
\node[figtext] at (4.6,.66) {$\Pi[\ell,1/2]$\\$\ell_r>0$};
\node[figtext] at (6.35,1.62) {Pooling\\$\delta_\pi$};
\path (0,{4/3})--(2,2) node[pos=.38,sloped,below=3pt,font=\footnotesize] {$3r=\kappa+4$};
\path (2,2)--(4,0) node[pos=.52,sloped,above=3pt,font=\footnotesize] {$r=4-\kappa$};
\path (2,2)--(7.3,0) node[pos=.54,sloped,above=4pt,font=\footnotesize] {$\kappa=r+(1-r/2)/D^2$};
\fill (2,2) circle (2pt);
\end{tikzpicture}
\caption{Optimal public supports with a common threshold, for a fixed $\pi\in(0,1/2)$ and $D=1/2-\pi$. The labelled straight-line boundaries are exact; the horizontal spacing of the symbolic intercept $D^{-2}>4$ is schematic. The upper-left region has $y>1/2$ and public information expands with $r$; the middle wedge has $0<\ell<\pi$ and public information contracts. On the lower-contact boundaries, $\ell=0$ or $\ell=\pi$, respectively. At the marked corner $(\kappa,r)=(2,2)$, every mean-$\pi$ law on $[0,1/2]$ is optimal.}
\label{fig:benchmark-regions}
\end{figure}
\FloatBarrier

\section{Dispersed decision thresholds}\label{sec:heterogeneity}
We now allow receivers to differ in their decision thresholds. The same public posterior can then induce some receivers to accept and others to reject. Verification can overturn the former decisions and reverse the latter, so both effects operate after the same public news. We show that the boundary at $\kappa=2$ in the common-threshold benchmark does not extend to this case.

\subsection{Cancellation and an exact smooth family}
If thresholds are uniform on $[0,1]$, then $Q(\mu)=\mu(1-\mu)$ and \eqref{eq:Hidentity} gives $H\equiv0$. Consequently $s(\mu)=\mu$ for every $r$. For $\kappa>0$ the unique optimum is pooling; for $\kappa=0$ every feasible public experiment is optimal. This cancellation is not specific to uniform inspection costs. For any common cost distribution independent of $t$, condition on a realized cost $c$. When $c\leq2\mu(1-\mu)$, the inspecting thresholds lie between
\[
L_c=\frac{c}{2(1-\mu)},\qquad U_c=1-\frac{c}{2\mu}.
\]
Acceptance is $(1-\mu)F(L_c)+\mu F(U_c)=\mu$ under a uniform $F$; if no threshold inspects, acceptance is again $\mu$. Averaging over $c$ preserves the equality.

To examine departures from this cancellation, consider
\begin{equation}\label{eq:family}
F_\alpha(t)=(1-\alpha)t+\alpha(3t^2-2t^3),\qquad 0\leq\alpha\leq1.
\end{equation}
The density is symmetric around $1/2$ and more concentrated there when $\alpha$ is larger. Indeed,
\[
\int_0^x[F_\alpha(t)-t]\dd t=-\frac\alpha2x^2(1-x)^2.
\]
Thus increasing $\alpha$ reduces threshold dispersion in the convex order. Direct integration yields
\begin{equation}\label{eq:s-family}
s(\mu)=(1-\alpha)\mu+
\alpha\{3\mu^2-2\mu^3+r\mu^2(1-\mu)^2(1-2\mu)\}.
\end{equation}
For $\alpha>0$, only the ratio $K=\kappa/\alpha$ matters for the optimal experiment. The linear term $(1-\alpha)\mu$ has the same expectation under every feasible experiment.

Define
\begin{align}\label{eq:chi}
\chi(y)&=(1-y)(8y^2-7y+1),&
T(y)&=3-K-4y+r\chi(y),\\
u&=\frac{7-\sqrt{17}}{16},&v_*&=\frac{7+\sqrt{17}}{16},\nonumber\\
K_-&=\frac{5-\sqrt{17}}4,&K_+&=\frac{5+\sqrt{17}}4.\nonumber
\end{align}
Here $u$ and $v_*$ are the two interior roots of $\chi$. We reserve $v(\mu)$ for the adjusted sender payoff.

\Needspace{7\baselineskip}
\begin{proposition}\label{prop:smooth}
Let $F=F_\alpha$ with $\alpha>0$ and $K=\kappa/\alpha$. Put $K_P=3-4\pi+r\chi(\pi)$.
\begin{enumerate}[label=(\roman*)]
\item If $K<K_P$, the unique optimal experiment is $\Pi[0,y]$, where $y>\pi$ is the unique solution of $T(y)=0$. If $K\geq K_P$, the unique optimum is pooling.
\item On the informative branch, cheaper verification raises $y$ when $K<K_-$ or $K>K_+$, and lowers $y$ when $K_-<K<K_+$. At $K=K_-$ and $K=K_+$ the contact is respectively $v_*$ and $u$, whenever it is above the prior. The region $K>K_+$ can be informative only if $\pi<u$.
\item Increasing $\kappa$ weakly contracts public information. At a fixed positive $\kappa$, increasing $\alpha$ weakly expands it. Both comparisons are strict on the informative branch.
\item On the informative branch, sender value rises with $r$ if and only if $K>1-r/4$, and falls if $K<1-r/4$. Under pooling its derivative with respect to $r$ is strictly positive.
\end{enumerate}
\end{proposition}

To see how the optimal experiment is obtained, remove the linear term from the sender's payoff and divide by $\alpha$. This gives
\begin{equation}\label{eq:reducedv}
\widetilde v(\mu)=(3-K+r)\mu^2-(2+4r)\mu^3+5r\mu^4-2r\mu^5.
\end{equation}
The derivative of $\widetilde v(\mu)/\mu$ is $T(\mu)$, and
\begin{equation}\label{eq:Tprime}
T'(\mu)=-4+r(-8+30\mu-24\mu^2)\leq-5/4.
\end{equation}
The ratio therefore has at most one interior maximizer. When it exists, the line from the origin to that point lies above the payoff everywhere, and the payoff is strictly concave to its right. These observations, together with the case in which the ratio is decreasing throughout, establish both the global optimality of the informative experiment and the pooling region.

The sender benefits from cheaper verification when the positive posterior is below $1/2$. The envelope derivative is
\begin{equation}\label{eq:Vr-smooth}
V_r=\alpha\pi J(y),\qquad J(y)=y(1-y)^2(1-2y).
\end{equation}
Because $T(1/2)=1-K-r/4$, this is exactly the condition in Proposition~\ref{prop:smooth}(iv). In particular,
\begin{equation}\label{eq:contraction-gain}
1-r/4<K<\min\{K_+,K_P\}
\end{equation}
gives an informative experiment that contracts with cheaper verification while the sender gains. This interval is nonempty: $K_P>1-r/4$ follows from strict monotonicity of $T$ and $\pi<1/2$.

There is also an additional expansion region absent from the common-threshold cost classification. If $\pi<u$ and
\begin{equation}\label{eq:onset}
3-4\pi<K<3-4\pi+2\chi(\pi),
\end{equation}
the sender pools until $r=[K-(3-4\pi)]/\chi(\pi)$ and discloses information above this threshold. These values satisfy $K>K_+$. Cheaper verification can therefore make public information worth providing even when its effective cost is high.

Figure~\ref{fig:smooth-regions} displays all three responses for the prior range $\pi<u$. The two horizontal boundaries fix the posterior contact, whereas the sloping boundary marks the point at which that contact reaches the prior. For $\pi\geq u$, the upper expansion region is absent: its contact could not lie above the prior.

\begin{figure}[t]
\centering
% Symbolic vertical coordinates preserve the exact order of all boundaries.
\begin{tikzpicture}[x=5cm,y=1.0cm,font=\small]
\fill[figblue!12] (0,0) rectangle (2,.65);
\fill[figwarm!12] (0,.65) rectangle (2,3.0);
\fill[figblue!12] (0,3.0)--(2,3.0)--(2,4.55)--(0,3.65)--cycle;
\fill[black!10] (0,3.65)--(2,4.55)--(2,5.25)--(0,5.25)--cycle;
\draw[boundary] (0,.65)--(2,.65);
\draw[boundary] (0,3.0)--(2,3.0);
\draw[boundary] (0,3.65)--(2,4.55);
\draw[guide] (2,0)--(2,5.25);
\draw[figaxis] (0,0)--(2.16,0) node[right] {$r$};
\draw[figaxis] (0,0)--(0,5.55) node[above] {$K=\kappa/\alpha$};
\foreach \x/\lab in {0/0,2/2}
 {\draw (\x,0)--(\x,-.08);\node[below] at (\x,-.1) {$\lab$};}
\foreach \y/\lab in {.65/K_-,3.0/K_+,3.65/{3-4\pi}}
 {\draw (0,\y)--(-.035,\y);\node[left] at (-.045,\y) {$\lab$};}
\node[figtext,text=figblue] at (1,.30) {Expansion: $y_r>0$};
\node[figtext] at (1,1.85) {Contraction: $y_r<0$\\$\Pi[0,y]$, $\pi<y<1$};
\node[figtext,text=figblue] at (1.2,3.43) {Expansion: $y_r>0$\\$\pi<y<u$};
\node[figtext] at (.53,4.83) {Pooling: $\delta_\pi$};
\node[right,font=\footnotesize] at (2.03,.65) {$y=v_*$};
\node[right,font=\footnotesize] at (2.03,3.0) {$y=u$};
\path (0,3.65)--(2,4.55) node[pos=.72,sloped,above=4pt,font=\footnotesize] {$K=K_P$};
\end{tikzpicture}
\caption{The smooth-threshold disclosure regions when $0<\pi<u$ and $\alpha>0$. The exact boundaries are $K_-=(5-\sqrt{17})/4$, $K_+=(5+\sqrt{17})/4$, and $K=K_P=3-4\pi+r\chi(\pi)$. Their vertical spacing is schematic. Since $\chi(\pi)>0$ in this prior range, the pooling boundary rises with $r$, and $K_P(0)>K_+$. At $K_-$ and $K_+$ the informative contact is constant in $r$; at $K=K_P$ the unique optimum is pooling.}
\label{fig:smooth-regions}
\end{figure}

The effect of threshold concentration in Proposition~\ref{prop:smooth} follows from the form of \eqref{eq:family}. Increasing $\alpha$ scales up the nonlinear part of the sender's payoff and hence lowers the effective information cost $\kappa/\alpha$. An arbitrary contraction of the threshold distribution need not change the payoff in this way. We also examine small departures from the common threshold in Appendix~\ref{app:concentration}. The benchmark's optimal experiment in its active expansion region is preserved exactly for all sufficiently concentrated atomless distributions with mean $1/2$. Its active contraction response is preserved for a uniform band of thresholds.
\FloatBarrier

\Needspace{22\baselineskip}
\subsection{Cheaper verification can leave every type worse informed}\label{sec:type-loss}
We are now in a position to compare the information receivers ultimately obtain and the welfare they derive from it. The following result allows any atomless threshold distribution sufficiently close to the common-threshold benchmark. It uses a different parameter region from the comparison in which cheaper verification benefits the sender.

\begin{proposition}\label{prop:type-loss}
Fix $\pi\in(0,1/2)$ and choose
\begin{equation}\label{eq:loss-parameters}
2<\kappa<4(1-\pi),\qquad 4\pi<r_1<4-\kappa.
\end{equation}
There exists $\bar\delta>0$ such that, for every atomless $F$ with mean $1/2$ supported on $[1/2-\delta,1/2+\delta]$, where $0<\delta<\bar\delta$, the following statements hold when verification becomes cheaper from $r_1$ to $r_2=2$.
\begin{enumerate}[label=(\roman*)]
\item Every initial optimal experiment assigns positive probability to zero and otherwise has support in the threshold band. The unique final optimal experiment is pooling.
\item For every threshold in the support of $F$, inspection becomes more frequent, the final experiment is a strict Blackwell garbling of its initial counterpart, and total expected loss increases. These statements average over the costs of that threshold type and hold for every initial optimal experiment.
\item Inspection expenditure decreases for every threshold type, and optimized sender value decreases.
\end{enumerate}
The same strict comparisons hold for $r_2<2$ sufficiently close to $2$.
\end{proposition}

The restrictions in \eqref{eq:loss-parameters} define a nonempty set for every prior under consideration. The mechanism is easiest to see as thresholds approach $1/2$. The initial experiment converges to $\Pi[0,1/2]$, while the final experiment is pooling. Inspection probabilities converge to $r_1\pi$ and $2\pi$, respectively. Yet the mass on posterior one falls from $r_1\pi/2$ to $2\pi^2$, and the mass on posterior zero also falls. Receivers inspect more often but are less likely to learn either state conclusively, because the public signal no longer contributes information. These strict inequalities hold uniformly for small threshold dispersion and, together with the location of the remaining posteriors, establish the Blackwell ordering.

For each threshold type, the limiting increase in total loss is
\begin{equation}\label{eq:loss-margin}
L_2-L_1\longrightarrow\pi\left(\frac{r_1}4-\pi\right)>0.
\end{equation}
Expected inspection expenditure falls by the same amount in the limit. The increase in decision loss is twice the increase in total loss. Thus the receivers' more frequent checks do not compensate for the disappearance of informative public news, despite being cheaper in expenditure terms. The sender also loses:
\begin{equation}\label{eq:sender-loss-margin}
V_1-V_2\longrightarrow
\pi\left[\frac{4-r_1-\kappa}{2}+(\kappa-2)\pi\right]>0.
\end{equation}
Thus cheaper verification can leave both the sender and every threshold type worse off. The sender gains identified in Propositions~\ref{prop:benchmark} and~\ref{prop:smooth} occur in other parameter regions.

Figure~\ref{fig:call-gap} illustrates the information comparison. Initially, the final posterior law has more mass at both fully revealing posteriors. Every remaining nonzero posterior in that law lies above $\pi$. The endpoint inequalities therefore imply that the initial call transform is strictly greater at every interior cutoff, establishing that the final experiment becomes less informative.

\begin{figure}[t]
\centering
\begin{tikzpicture}[x=1.08cm,y=2.5cm,font=\small]
\fill[black!7] (4.4,0) rectangle (5.6,1.55);
\draw[figaxis] (0,0)--(10.35,0) node[right] {$x$};
\draw[figaxis] (0,0)--(0,1.74) node[above] {$\Delta C_t(x)$};
\draw[guide] (1.7,0)--(1.7,1.36);
\draw[guide] (4.4,0)--(4.4,.65636);
\draw[guide] (5.6,0)--(5.6,.44);
\draw[figblue,thick,dashed] (1.7,.83)--(10,0);
\draw[payoff,thick] (0,0)--(1.7,1.36)--(4.4,.65636)
 .. controls (4.8,.55212) and (5.2,.48) .. (5.6,.44)--(10,0);
\foreach \x/\lab in {0/0,1.7/\pi,4.4/a,5.6/b,10/1}
 {\draw (\x,0)--(\x,-.035);\node[below] at (\x,-.045) {$\lab$};}
\node[figtext] at (3.5,1.60) {$C_{\Xi_{t,1}}-C_{\Xi_{t,2}}$};
\draw[thin,-{Stealth[length=3pt]}] (3.5,1.43)--(3.4,.93);
\node[figtext] at (.56,1.06) {$x\Delta P_0$};
\node[figtext,text=figblue] at (7.9,.92) {$\Delta A(1-x)$\\lower bound for $x\geq\pi$};
\draw[figblue,thin,-{Stealth[length=3pt]}] (7.5,.71)--(7.15,.285);
\node[font=\footnotesize,align=center] at (5,1.35) {threshold\\band};
\end{tikzpicture}
\caption{A typewise Blackwell comparison in Proposition~\ref{prop:type-loss}. For each threshold type, put $\Delta A=A_1-A_2>0$ and $\Delta P_0=P_{01}-P_{02}>0$, where $A_j$ and $P_{0j}$ are its final masses at one and zero. The exact call-transform gap is $x\Delta P_0$ on $[0,\pi]$, is at least $\Delta A(1-x)$ on $[\pi,1]$, and equals this lower bound for $x\geq b$. Here $a=1/2-\delta$ and $b=1/2+\delta$; heights and the bend inside the band are schematic. The endpoint inequalities follow from $A_1\to r_1\pi/2$, $A_2\to2\pi^2$, $P_{01}\to1-2\pi+r_1\pi/2$, and $P_{02}\to2\pi(1-\pi)$, with $r_1>4\pi$.}
\label{fig:call-gap}
\end{figure}
\FloatBarrier

\section{Shared verification}\label{sec:sharing}
We now allow a receiver to benefit from someone else's inspection. If receiver $i$ inspects, its finding is automatically broadcast to every other receiver with probability $\rho\in[0,1]$. Broadcast draws are independent across inspectors and independent of the state and types. An inspector always learns its own finding. Inspection decisions are simultaneous, so a receiver cannot wait for another's result before deciding whether to pay. Broadcasts arrive before actions are chosen, and any broadcast reveals the state perfectly.\footnote{This is a technology of automatic dissemination. It does not give an inspector a strategic choice about whether to conceal an unfavorable finding. Such a choice would introduce another signaling problem.}

Throughout this section $F$ is atomless. At an interior public posterior, set $a=rQ(\mu)<1$ and let $p_i$ be receiver $i$'s inspection probability. If no other receiver broadcasts, inspection is worth $q(t_i,\mu)$; otherwise it is redundant. Hence
\begin{equation}\label{eq:BNE}
p_i=a\prod_{j\ne i}(1-\rho p_j).
\end{equation}
This system has a unique equilibrium. In particular, subtracting the equations for $i$ and $j$ gives
\[
(p_i-p_j)\left[1-a\rho\prod_{k\ne i,j}(1-\rho p_k)\right]=0.
\]
The bracket is strictly positive, so every equilibrium is symmetric. The common probability is the unique solution of
\begin{equation}\label{eq:p-scalar}
p=a(1-\rho p)^{n-1},\qquad h=(1-\rho p)^{n-1}=p/a.
\end{equation}
Here $h$ is the probability that no other receiver broadcasts. The cost cutoff of threshold type $t$ is $q(t,\mu)h$. At $\rho=0$, $h=1$ and $p=rQ$, recovering independent verification.

\subsection{Free riding, final information, and payoffs}\label{sec:sharing-reduction}
The probability that threshold type $t$ ends up knowing the state is
\begin{equation}\label{eq:lambda-sharing}
\lambda_t=1-h+rq(t,\mu)h^2.
\end{equation}
The first term reflects a broadcast by another receiver. In the absence of such a broadcast, own inspection occurs with probability $rq(t,\mu)h$. Replacing $rq$ by $\lambda_t$ in \eqref{eq:finalkernel} gives the final posterior kernel under sharing.

The inspection probability, expenditure, decision loss, and total loss of a threshold type are
\begin{equation}\label{eq:sharing-welfare}
\Lambda_t=rqh,\qquad E_t=\frac r2q^2h^2,\qquad
D_t=qh-rq^2h^2,\qquad L_t=qh-\frac r2q^2h^2.
\end{equation}
For a fixed realized cost, expected loss is simply $\min\{c,qh\}$. Thus any reduction in $h$ weakly improves the welfare of every fixed-cost receiver. Under the uniform cost distribution, the improvement in average welfare is strict whenever $h$ falls and $q>0$.

Sharing lowers $h$ and the inspection probability $p$. With $m=n-1$ and $d=1+(n-2)\rho p$, implicit differentiation yields
\begin{equation}\label{eq:pderivatives}
p_a=\frac{h(1-\rho p)}d>0,\qquad
p_\rho=-\frac{mp^2}d<0.
\end{equation}
Wider sharing therefore reduces private inspection. To determine whether receivers nevertheless learn more from the available checks, average \eqref{eq:lambda-sharing} over thresholds:
\begin{equation}\label{eq:lambda-avg}
\lambda=1-h+ah^2=1-h+hp,\qquad
\lambda_\rho=(2p-1)h_\rho.
\end{equation}
When inspection is sufficiently likely, $p>1/2$, the reduction in own inspection outweighs the gain from dissemination, and aggregate final information falls. Welfare nevertheless improves: the prospect of obtaining someone else's finding lowers the expected loss from remaining uninformed without paying for one's own check, as expressed by $\min\{c,qh\}$.

For the sender, write $D_F(\mu)=F(\mu)-\mu$. Its expected acceptance becomes
\begin{equation}\label{eq:s-sharing}
s(\mu;r,\rho,n)=\mu+hD_F(\mu)+rh^2H(\mu).
\end{equation}
Uniform thresholds again imply exact cancellation, because both $D_F$ and $H$ vanish. Under \eqref{eq:family}, let $z=\mu(1-\mu)$ and $w=1-2\mu$. Then
\begin{equation}\label{eq:sharing-family}
Q=z(1+\alpha z),\qquad D_F=-\alpha zw,\qquad H=\alpha z^2w,
\end{equation}
and
\begin{equation}\label{eq:s-sharing-family}
s=\mu-\alpha zw h(1-rzh).
\end{equation}
Since $rzh\leq1/2$, wider sharing raises acceptance at every fixed posterior below $1/2$ and lowers it above $1/2$. This pointwise payoff comparison does not determine how the optimal public experiment changes.

\subsection{Cheaper inspection and wider sharing can move disclosure differently}
We next compare cheaper inspection with wider sharing when the sender can redesign its experiment. We consider sufficiently small sharing probabilities and establish optimality over all feasible public experiments.

\begin{proposition}\label{prop:opposite}
Fix $n\geq2$, $\alpha\in(0,1]$, and $\pi<u$, where $u$ is defined in \eqref{eq:chi}. Let $z_u=u(1-u)$ and choose
\begin{equation}\label{eq:opposite-r}
0<r<\frac{\alpha}{2+4\alpha z_u}.
\end{equation}
There exists $\varepsilon>0$ such that, for every fixed $K=\kappa/\alpha\in(K_+-\varepsilon,K_+)$, there is $\bar\rho(K)>0$ for which the following holds. For all $\rho\in(0,\bar\rho(K))$, the sender's unique optimal experiment is $\Pi[0,y]$, with $\pi<y<1/2$, and
\begin{equation}\label{eq:opposite-signs}
y_r<0<y_\rho,\qquad V_r>0,\qquad V_\rho>0.
\end{equation}
Thus cheaper inspection contracts public information, while wider sharing expands it, and both changes benefit the sender.
\end{proposition}

At the boundary $K=K_+$ and $\rho=0$, the optimal contact is $u$ and is locally invariant to $r$. Moving $K$ just below that boundary makes the response to $r$ negative. The response to sharing is governed by a different margin. At $\rho=0$, the effect of sharing on the derivative of the relevant payoff ratio, evaluated at $u$, is
\begin{equation}\label{eq:sharing-contact}
(n-1)rJ(u)(1-2u)\bigl[\alpha-2r-4\alpha rz_u\bigr]>0.
\end{equation}
The restriction \eqref{eq:opposite-r} gives the strict inequality. Because the ratio derivative is decreasing in its posterior argument, a positive shift moves the contact to the right. Appendix~\ref{app:opposite} proves that the supporting line remains globally optimal for sufficiently small sharing.

Both interventions increase the sender's payoff because the positive public posterior remains below $1/2$, where their effect on aggregate acceptance is favorable. They change acquisition incentives differently, however. At a fixed posterior, cheaper inspection raises the probability of one's own check, whereas wider sharing reduces it. The sender's disclosure choice depends on how these effects vary across public posteriors. This is why the same sign of the payoff effect can be accompanied by opposite changes in the optimal experiment. The comparison holds for the prior range and small sharing probabilities specified in Proposition~\ref{prop:opposite}.

\subsection{Wider sharing can reduce final information while improving welfare}
\eqref{eq:lambda-avg} shows how sharing can reduce final information at a fixed public posterior. To carry this comparison through to the full game, we also need to account for the sender's choice. We do so by identifying a region in which pooling is optimal at every sharing probability.

\begin{proposition}\label{prop:sharing-loss}
\begin{enumerate}[label=(\roman*)]
\item At a fixed interior public posterior, increasing $\rho$ strictly improves the cost-averaged welfare of each threshold type and lowers the equilibrium inspection probability. Aggregate final information decreases locally exactly when $p>1/2$.
\item Suppose $n=2$, $F=F_\alpha$ with $\alpha>0$, and $\kappa>19\alpha$. The unique optimal public experiment is pooling for every $r\in(0,2]$ and $\rho\in[0,1]$. If
\begin{equation}\label{eq:golden}
a_\pi:=r\pi(1-\pi)[1+\alpha\pi(1-\pi)]>\frac{\sqrt5-1}{2},
\end{equation}
increasing $\rho$ from zero to one strictly reduces aggregate final information in the Blackwell order, lowers inspection, and strictly improves every threshold type's cost-averaged welfare. The parameter region in \eqref{eq:golden} is nonempty.
\end{enumerate}
\end{proposition}

For two receivers, $h=(1+\rho a)^{-1}$ and $p=a/(1+\rho a)$. Under full sharing the aggregate probability of learning the state is $1-(1+a)^{-2}$, whereas without sharing it is $a$. Their difference is
\begin{equation}\label{eq:lambda-golden}
\left[1-\frac1{(1+a)^2}\right]-a
=\frac{a(1-a-a^2)}{(1+a)^2}.
\end{equation}
This expression is negative precisely under \eqref{eq:golden}. Because the public posterior is unchanged, a reduction in the probability of replacing it by a fully revealing posterior is an exact Blackwell contraction of the aggregate final experiment. The sufficient condition $\kappa>19\alpha$ is a uniform curvature bound, not a sharp cost threshold.

To understand the welfare gain, recall that the aggregate posterior law suppresses preference labels. Its contraction need not make the final experiment less informative for every threshold type. In addition, receivers pay for their own acquisition. Wider sharing reduces the need to pay for a check, and every threshold type benefits after averaging over its private costs. The aggregate information comparison alone therefore does not determine welfare.

Figure~\ref{fig:sharing-regions} distinguishes the local and endpoint comparisons. With two receivers, $p>1/2$ is equivalent to $a>1/(2-\rho)$, which gives the curved boundary for the local effect of sharing. The horizontal boundary instead compares $\rho=0$ with $\rho=1$. Final information can fall initially and then recover, yet remain below its no-sharing level. Throughout, wider sharing reduces inspection and improves cost-averaged welfare at the fixed public posterior.

\begin{figure}[t]
\centering
\begin{tikzpicture}[x=9.0cm,y=7.0cm,font=\small]
\pgfmathsetmacro{\goldenbound}{(sqrt(5)-1)/2}
\fill[figblue!9] (0,0) rectangle (1,1);
\fill[figwarm!16] (0,1)--(1,1)
 --plot[domain=1:0,samples=81] (\x,{1/(2-\x)})--cycle;
\draw[guide] (0,1)--(1,1)--(1,0);
\draw[boundary,domain=0:1,samples=81] plot (\x,{1/(2-\x)});
\draw[black!60,dashed] (0,\goldenbound)--(1,\goldenbound);
\draw[figaxis] (0,0)--(1.08,0) node[right] {$\rho$};
\draw[figaxis] (0,0)--(0,1.1) node[above] {$a=rQ(\mu)$};
\foreach \x/\lab in {0/0,1/1}
 {\draw (\x,0)--(\x,-.016);\node[below] at (\x,-.025) {$\lab$};}
\foreach \y/\lab in {.5/{1/2},1/1}
 {\draw (0,\y)--(-.012,\y);\node[left] at (-.018,\y) {$\lab$};}
\node[right] at (1.015,\goldenbound) {$\dfrac{\sqrt5-1}{2}$};
\node[figtext] at (.23,.84) {$\lambda_\rho<0$\\$p>1/2$};
\node[figtext,text=figblue] at (.58,.30) {$\lambda_\rho>0$\\$p<1/2$};
\node[figtext] at (.68,.98) {$a=\dfrac1{2-\rho}$};
\draw[thin,-{Stealth[length=3pt]}] (.78,.91)--(.81,{1/1.19});
\end{tikzpicture}
\caption{Sharing and aggregate final information for $n=2$ at a fixed interior public posterior, with $0<a<1$. The curved boundary is the exact locus $p=1/2$: above it, wider sharing reduces the revelation probability $\lambda$; below it, wider sharing increases $\lambda$. The dashed horizontal line gives the distinct endpoint comparison: $\lambda(1)<\lambda(0)$ precisely when $a>(\sqrt5-1)/2$. Both sides of the curved boundary have lower inspection and higher cost-averaged receiver welfare as sharing rises. The diagram applies to the continuation game for any atomless $F$. Proposition~\ref{prop:sharing-loss}(ii) supplies the complete-game application by ensuring optimal pooling and evaluating $a$ at $a_\pi$.}
\label{fig:sharing-regions}
\end{figure}
\FloatBarrier

\section{Discussion}\label{sec:discussion}
We have studied how a sender's public disclosure responds to receivers' ability to verify the state. Verification can undermine acceptance after favorable public news, but it can also generate acceptance after unfavorable news. When public information is costly, the sender may benefit from cheaper verification and provide less information of its own. The distribution of receivers' decision thresholds determines how these two effects combine. Sharing introduces a further response: receivers inspect less when they can learn from one another, and the sender may then disclose more.

These responses matter for evaluating measures that make claims easier to check or findings easier to share. Cheaper verification can increase inspection while leaving every threshold type worse informed and worse off. In a different comparison, wider sharing can reduce aggregate final information while improving every threshold type's welfare. The first effect follows from the withdrawal of public information; the second combines a change in the allocation of information with lower acquisition costs. Neither inspection frequency nor the informativeness of the public experiment is sufficient to assess the welfare consequences.

The exact results use a discrete verification technology and uniformly distributed private inspection costs. Our treatment of heterogeneous preferences gives a global characterization within a smooth family of threshold distributions, together with robustness results near a common threshold. It does not imply the same comparative statics for arbitrary changes in preferences. Likewise, the opposite disclosure responses to costs and sharing are established for small sharing probabilities, and the full-game information and welfare comparison under sharing uses a sufficient condition for pooling. Extending these comparisons to more general acquisition technologies and cost distributions would help determine how far the mechanisms extend beyond the cases analyzed here.

The form of dissemination also matters. We assume that a shared finding reaches the entire audience. With a fixed positive probability of such a broadcast, a large audience can obtain information at little expected cost to each receiver. Appendix~\ref{app:large} shows that, for a smooth threshold distribution and any fixed $\kappa>0$, the sender eventually provides no public information, while receivers approach full information and zero expected loss. This conclusion relies on average sender stakes and broadcasts to the whole audience. A bounded number of contacts, a transmission probability that falls with audience size, or a positive lower bound on inspection costs can change the comparison. These are natural settings in which to study the interaction between public disclosure and the information that receivers supply to one another.

\appendix
\section{Preliminaries}\label{app:prelim}
\subsection{Implementation and supporting lines}
Every law $\Pi$ on $[0,1]$ satisfying \eqref{eq:bayes} can be implemented by a signal whose realization is its posterior. For any Borel set $B$, define the state-contingent signal laws by
\[
\Prb(\mu\in B\mid\theta=1)=\frac1\pi\int_B\mu\dd\Pi(\mu),\qquad
\Prb(\mu\in B\mid\theta=0)=\frac1{1-\pi}\int_B(1-\mu)\dd\Pi(\mu).
\]
Bayes' rule gives posterior $\mu$ almost surely. The expected quadratic information cost equals $\kappa\E\mu^2-\kappa\pi^2$, which proves \eqref{eq:sender} by the usual concavification argument.

We repeatedly use the following supporting-line observation. If an affine function $L$ majorizes $v$ everywhere and a feasible experiment is supported on the contact set $\{\mu:L(\mu)=v(\mu)\}$, that experiment is optimal. Conversely, every optimal experiment must be supported on that contact set whenever the majorant attains the concavified value at the prior. If there are precisely two contacts bracketing $\pi$, Bayes plausibility uniquely determines the probabilities. A strict tangent at $\pi$ establishes unique pooling.

\subsection{Binary experiments and final information}
For binary states, the posterior representation identifies Blackwell dominance with convex order among laws of the same mean. On $[0,1]$, every convex function is an affine function plus an integral of the functions $(\mu-x)_+$. This proves the call-transform criterion \eqref{eq:call}. For a two-point law $\Pi[a,b]$, replacing either endpoint by a more extreme posterior while maintaining the mean is a mean-preserving spread. This proves the interval-inclusion comparison used in the text.

Equation \eqref{eq:finalkernel} follows because the event of inspection is conditionally independent of the state given the public posterior and types. The same is true of the event of receiving a broadcast under the state-independent dissemination technology. Thus the absence of an inspection or broadcast does not itself change the public posterior. Averaging the resulting posterior laws over independently drawn costs and thresholds preserves their mean. Direct integration of $c$ over $[0,q]$ gives expected expenditure $rq^2/2$, establishing \eqref{eq:ind-welfare}.

\section{Proof of Proposition~\ref{prop:benchmark}}\label{app:benchmark}
Write the two branches of the adjusted payoff as
\[
v_-(\mu)=(r-\kappa)\mu^2,\qquad
v_+(\mu)=1-r+2r\mu-(r+\kappa)\mu^2.
\]
The lower branch applies below $1/2$ and the upper branch at and above $1/2$. Their jump is $1-r/2\geq0$.

\subsection{The region \texorpdfstring{$\kappa<2$}{kappa < 2}}
The chord from zero to $(1/2,v_+(1/2))$ has slope $m_0=2-(r+\kappa)/2>0$. It majorizes the lower branch. Indeed, if $r\leq\kappa$ this is immediate; otherwise $v_-(\mu)/\mu\leq(r-\kappa)/2\leq m_0$, since $r\leq2$. The chord majorizes the upper branch precisely when
\[
m_0\geq v_+'(1/2)=r-\kappa,
\]
or $3r\leq\kappa+4$. The contact set is then $\{0,1/2\}$, apart from the excluded corner $r=\kappa=2$.

If $3r>\kappa+4$, then $r>1$ and $r>\kappa$. The tangent to $v_+$ passing through zero has contact
\[
y=\sqrt{\frac{r-1}{r+\kappa}}>\frac12.
\]
Concavity of $v_+$ makes this tangent a global majorant of the upper branch. At $1/2$ the gap from that branch is $(r+\kappa)(y-1/2)^2>0$. Since $v_-$ is convex and starts at zero, a line through zero above its value at $1/2$ strictly majorizes it on $(0,1/2]$. Hence the tangent majorizes the entire payoff, with contacts only at zero and $y$. This proves the unique optimum.

In the active region,
\[
y_r=\frac{1+\kappa}{2y(r+\kappa)^2}>0.
\]
The envelope derivative on both branches of the solution is
\[
V_r=-\frac{\pi(1-y)^2}{y}<0.
\]
At $r=\kappa=0$, the payoff is zero below $1/2$ and one above it; the same chord argument gives the unique law $\Pi[0,1/2]$ directly.

\subsection{The region \texorpdfstring{$\kappa>2$}{kappa > 2}}
Here $v_-$ is strictly concave. A tangent at $x<1/2$ meets the upper branch at $1/2$ if and only if
\begin{equation}\label{eq:benchmark-tangent}
(\kappa-r)(1/2-x)^2=1-r/2.
\end{equation}
The solution is $x=1/2-\sqrt{(1-r/2)/(\kappa-r)}$. Whenever $0<x<\pi$, this tangent strictly majorizes $v_-$ except at $x$. At the upper contact its slope exceeds $v_+'(1/2)$ by $(\kappa-r)(1-2x)>0$. Concavity of $v_+$ therefore makes it a strict majorant to the right of $1/2$. Its only contacts are $x$ and $1/2$.

If $x\leq0$, equivalently $r+\kappa\leq4$, the chord from zero to the upper branch at $1/2$ has nonnegative slope and lies above the nonpositive lower branch. Its slope also exceeds the upper derivative, because $\kappa>2$ and $r+\kappa\leq4$ imply $\kappa+4\geq3r$. Hence $\Pi[0,1/2]$ is optimal. If $x\geq\pi$, the tangent to $v_-$ at $\pi$ majorizes the upper branch: tangent minus payoff at $1/2$ equals
\[
(\kappa-r)(1/2-\pi)^2-(1-r/2)\geq0,
\]
and the tangent slope exceeds the upper derivative. It is therefore a global majorant. The only contact compatible with mean $\pi$ is $\pi$ itself. This also covers $r=2$, for which the untruncated $x$ equals $1/2$.

The interior contact rises with $r$:
\begin{equation}\label{eq:ellr}
\ell_r=\frac{\kappa-2}{4(\kappa-r)^2(1/2-\ell)}>0.
\end{equation}
The boundaries of the active region are $r=4-\kappa$ and
\[
r_P=\frac{1-\kappa D^2}{1/2-D^2},
\]
whenever they lie in $[0,2]$. The envelope derivative is
\[
V_r=\frac{D\ell^2-(\pi-\ell)/4}{1/2-\ell}.
\]
Its numerator is strictly increasing in $\ell\geq0$, is negative at zero, and is positive at $\pi$. Its unique root is \eqref{eq:ellcrit}. Under pooling $V_r=\pi^2$. These facts establish all the value comparisons in part (ii).

If $\kappa=2$ and $r<2$, the chord at $1/2$ is a strict majorant away from zero and $1/2$. At $\kappa=r=2$, the adjusted payoff is zero on $[0,1/2]$ and equals $-(2\mu-1)^2$ above it. Precisely the mean-$\pi$ laws supported on $[0,1/2]$ are then optimal. This proves part (iii).

\subsection{Cost and welfare comparisons used in the text}
For $\Pi[0,y]$, expected inspection and total loss are
\[
\Lambda=\frac{\pi r(1-y)}y,\qquad
L=\frac\pi y\left[(1-y)-\frac r2(1-y)^2\right].
\]
The first decreases with $y$, while
\[
L_y=\pi\left[\frac{r/2-1}{y^2}-\frac r2\right]<0.
\]
Thus a rise in $\kappa$ that lowers $y$ increases both inspection and total loss. For $\Pi[\ell,1/2]$, every posterior is weakly below the threshold, so $\E q=\pi$ and
\[
\E q^2=\frac\pi2-D\ell,
\qquad \Lambda=r\pi,\qquad
L=\pi-\frac r2\left(\frac\pi2-D\ell\right).
\]
The rise in $\ell$ induced by a higher $\kappa$ therefore increases loss while leaving inspection unchanged. Constant-policy regions give weak comparisons; the expressions agree at their boundaries. The derivatives of $y$ and $\ell$ with respect to $\kappa$ stated in the text follow directly from their formulas.

Substitution of \eqref{eq:benchmark-tangent} into the objective gives \eqref{eq:premium}. Define
\[
A=\sqrt{1-r/2},\qquad h_0=\sqrt{\frac{1-r/2}{\kappa-r}}.
\]
In the interior region $0<h_0<1/2$ and $h_0>D$. The positive expression inside the square in \eqref{eq:premium} has derivative
\[
\frac{-1+2Dh_0}{4A}<0.
\]
This proves that the value of public persuasion above pooling falls with $r$. Setting $r=0$ in the support formulas gives the three cost regions and the values reported in the text. In that case all optimal posteriors are weakly below $1/2$, and $\E q=\pi$, proving the welfare statement.

\section{Proof of Proposition~\ref{prop:smooth}}\label{app:smooth}
For the density $f_\alpha=1-\alpha+6\alpha t(1-t)$, integration in \eqref{eq:QAH} gives
\[
Q=z(1+\alpha z),\quad H=\alpha z^2(1-2\mu),\quad
F_\alpha(\mu)-\mu=-\alpha z(1-2\mu).
\]
These identities prove \eqref{eq:s-family} and \eqref{eq:reducedv}. Let $g(\mu)=\widetilde v(\mu)/\mu$ for $\mu>0$, extended continuously at zero. Then $g'=T$. The quadratic $-8+30\mu-24\mu^2$ has maximum $11/8$ on $[0,1]$. Since $r\leq2$, \eqref{eq:Tprime} follows. Also $T(1)=-1-K<0$.

If $T(0)>0$, there is a unique root $y\in(0,1)$. The ratio $g$ increases strictly up to $y$ and decreases strictly thereafter. Consequently the line $\mu g(y)$ majorizes $\widetilde v$ globally and has exactly the contacts zero and $y$. To the right of $y$,
\[
\widetilde v''(\mu)=2T(\mu)+\mu T'(\mu)<0.
\]
The concave envelope is therefore the line on $[0,y]$ followed by $\widetilde v$ on $[y,1]$, with matching slopes at $y$. If $T(0)\leq0$, the same second-derivative identity makes the payoff strictly concave on $(0,1]$. These arguments prove informative disclosure precisely when $T(\pi)>0$, or $K<K_P$. At equality the prior is the contact itself, and Bayes plausibility forces pooling. This proves part (i), including uniqueness.

At an informative contact, implicit differentiation gives
\begin{equation}\label{eq:smooth-derivatives}
y_r=-\frac{\chi(y)}{T'(y)},\qquad y_K=\frac1{T'(y)}<0.
\end{equation}
The sign of $y_r$ is the sign of $\chi(y)$. This polynomial is positive below $u$ and above $v_*$, and negative between them. Since $T(u)=K_+-K$ and $T(v_*)=K_--K$, monotonicity of $T$ converts these posterior regions into the cost regions in part (ii). At the two boundaries the contacts are independent of $r$. If $K>K_+$, the root lies below $u$, so it can be informative only when $\pi<u$. Parts (ii) and (iii) follow, including the weak comparison across pooling boundaries.

The envelope derivative is the expectation of $H$. For $\Pi[0,y]$ it is $\alpha\pi J(y)$, proving \eqref{eq:Vr-smooth}. Its sign is positive exactly when $y<1/2$, which is equivalent to $K>1-r/4$ because $T$ is decreasing. Under pooling the derivative is $\alpha\pi^2(1-\pi)^2(1-2\pi)>0$. The onset calculation \eqref{eq:onset} follows by solving $T(\pi)=0$ for $r$ and using $\chi(\pi)>0$ when $\pi<u$.

\section{Robustness near a common threshold}\label{app:concentration}
The smooth family is not the only source of threshold robustness. This appendix treats distributions concentrated near $1/2$. It also provides the uniform bounds used in Proposition~\ref{prop:type-loss}.

\subsection{Exact preservation of the active expansion policy}
Fix a compact parameter set strictly inside
\[
\kappa<2,\qquad (\kappa+4)/3<r\leq2.
\]
The benchmark contact $y$ is uniformly above $1/2$. Let $F$ be any atomless distribution of mean $1/2$ supported on $[a,b]=[1/2-\delta,1/2+\delta]$. Below $a$, all receivers reject absent inspection, and their mean threshold gives $s(\mu)=r\mu^2$. Above $b$, all accept absent inspection, giving $s(\mu)=1-r(1-\mu)^2$. These identities are exact.

Inside the band, \eqref{eq:action} is bounded above by the acceptance payoff $1-rq(t,\mu)(1-\mu)$: the rejected receiver's acceptance probability is no larger because $rq\leq1$. Uniformly for $t,\mu$ in the band, $q(t,\mu)\to1/2$. Thus
\[
\sup_{\mu\in[a,b]}s(\mu)\leq1-r/4+o(1),
\]
where the error is uniform over the admissible distributions and the compact parameter set. The benchmark origin tangent has a strictly positive gap at $1/2$, namely $(r+\kappa)(y-1/2)^2$. For sufficiently small $\delta$, it therefore majorizes the adjusted payoff throughout the band strictly. Outside the band the benchmark branches are unchanged, so its only contacts remain zero and $y$. The exact policy $\Pi[0,y]$, and hence its strict expansion with $r$, survives.

\subsection{Preservation of the active contraction under a uniform band}
Now fix a compact subset of $\kappa>2$, $r<2$, and $0<\ell<\pi$ in the benchmark. Let $F$ be uniform on $[a,b]$ as above. Direct integration gives
\begin{equation}\label{eq:uniformband}
s(\mu)=\begin{cases}
r\mu^2,&\mu\leq a,\\
\dfrac{\mu-a+ra^2(1-2\mu)}{2\delta},&a\leq\mu\leq b,\\
1-r(1-\mu)^2,&\mu\geq b.
\end{cases}
\end{equation}
The tangent to $(r-\kappa)\mu^2$ at
\begin{equation}\label{eq:bandcontact}
x=b-\sqrt{\frac{1-r(a^2+b^2)}{\kappa-r}}
\end{equation}
passes through the adjusted payoff at $b$. For small $\delta$, it has $0<x<\pi<a$. Its gap from the lower branch is $(\kappa-r)(\mu-x)^2$. The middle branch of $s$ has slope $[1-2ra^2]/(2\delta)$, which tends uniformly to positive infinity because $r<2$ on the compact set. Hence the adjusted middle branch has slope above the tangent even at $b$. Their gap is positive to the left of $b$ and vanishes at $b$.

On the upper branch, the tangent slope exceeds the derivative at $b$ by
\[
2[(\kappa-r)(b-x)+r(b-a)]>0.
\]
Concavity of that branch proves strict majorization to the right. Thus the unique optimum is $\Pi[x,b]$. Differentiating \eqref{eq:bandcontact} gives
\[
x_r=\frac{\kappa(a^2+b^2)-1}{2(\kappa-r)^2(b-x)}>0
\]
for sufficiently small $\delta$. Public information contracts with cheaper verification. The sender's envelope derivative is
\[
V_r=\frac{(b-\pi)x^2-a^2(\pi-x)}{b-x}.
\]
It converges to the benchmark derivative uniformly on the compact set. Any strict sender-gain comparison in the interior benchmark therefore persists for small uniform threshold dispersion. This conclusion is local in dispersion and does not assert robustness under arbitrary large changes in $F$.

\section{Proof of Proposition~\ref{prop:type-loss}}\label{app:type-loss}
Fix parameters satisfying \eqref{eq:loss-parameters}. All limits in this proof are uniform over atomless mean-$1/2$ distributions supported on $[a,b]=[1/2-\delta,1/2+\delta]$, over thresholds in that band, and over initial optimal experiments. Take $\delta$ small enough that $\pi<a$.

\subsection{Initial and final optimal experiments}
At $r=r_1$, below the band the adjusted payoff is $(r_1-\kappa)\mu^2<0$, because $r_1<4-\kappa<\kappa$. At $b$, its ratio to $b$ converges to
\[
2-\frac{r_1+\kappa}{2}>0.
\]
Above the band, the derivative of the payoff ratio is
\[
\frac{r_1-1}{\mu^2}-(r_1+\kappa).
\]
It is negative for all $\mu\geq b$ when $\delta$ is small. If $r_1\leq1$ this is immediate. If $r_1>1$, its value near $1/2$ has sign $3r_1-\kappa-4<0$, since $r_1<4-\kappa$ and $\kappa>2$, and it decreases with $\mu$. The positive maximum of the payoff ratio is consequently attained in $[a,b]$. A line from zero with that slope majorizes the entire payoff; its contacts are zero and the ratio maximizers in the band. Any such contact can be combined with zero to attain mean $\pi$. Every optimal law is therefore supported on zero and the band. If $w$ is its probability on the band, Bayes plausibility implies
\[
\frac\pi b\leq w\leq\frac\pi a,
\]
so $w\to2\pi$ uniformly.

At $r_2=2$, below the band the adjusted payoff is $(2-\kappa)\mu^2$. Its tangent at $\pi$ has gap $(\kappa-2)(\mu-\pi)^2$ from that quadratic. Above the band, the actual adjusted payoff equals this quadratic minus $(2\mu-1)^2$, so the tangent strictly majorizes it. Within the band, each threshold type's acceptance probability converges uniformly to $1/2$. The adjusted payoff therefore converges uniformly there to $1/2-\kappa/4$. The tangent has a strictly positive limiting gap $(\kappa-2)(1/2-\pi)^2$ at the band. It is consequently a strict global majorant except at $\pi$ for small $\delta$. Pooling is uniquely optimal.

The same argument works for $r_2<2$ sufficiently close to $2$. On the lower branch the tangent gap remains $(\kappa-r_2)(\mu-\pi)^2$; the strictly positive gap away from $\pi$ persists uniformly. Near $\pi$ the payoff is exactly the same strictly concave quadratic. Thus unique pooling is preserved, not merely convergence of the optimal experiment.

\subsection{Final information}
For a fixed threshold $t$ in the band, let $A_j$ and $P_{0j}$ be the final masses on posteriors one and zero at $r_j$. Initially, $q(t,\mu)\to1/2$ uniformly on the nonzero public posteriors, and their total probability tends to $2\pi$. Hence
\[
A_1\to\frac{r_1\pi}{2},\qquad
P_{01}\to1-2\pi+\frac{r_1\pi}{2}.
\]
Under final pooling, $q(t,\pi)\to\pi$, so
\[
A_2\to2\pi^2,\qquad P_{02}\to2\pi(1-\pi).
\]
The first difference is positive by $r_1>4\pi$. For the second,
\[
1-2\pi+\frac{r_1\pi}{2}-2\pi(1-\pi)>(1-2\pi)^2>0.
\]
Thus $A_1>A_2$ and $P_{01}>P_{02}$ hold uniformly for small $\delta$.

Every nonzero initial posterior is either in $[a,b]$ or equals one, and hence exceeds $\pi$. For $0<x\leq\pi$, the call-transform difference is therefore
\[
C_{\Xi_{t,1}}(x)-C_{\Xi_{t,2}}(x)=x(P_{01}-P_{02})>0.
\]
For $\pi\leq x<1$, the final pooled experiment has call transform $A_2(1-x)$, whereas the initial transform is at least $A_1(1-x)$. This proves strict Blackwell dominance of the initial final-information law for every threshold type and every initial optimal experiment.

\subsection{Inspection, welfare, and sender value}
Using \eqref{eq:ind-welfare} and the same uniform limits gives
\begin{align*}
\Lambda_1&\to r_1\pi,&\Lambda_2&\to2\pi,\\
L_1&\to\pi-r_1\pi/4,&L_2&\to\pi-\pi^2,\\
E_1&\to r_1\pi/4,&E_2&\to\pi^2.
\end{align*}
All stated signs are strict under \eqref{eq:loss-parameters}. The decision-loss increase tends to $2\pi(r_1/4-\pi)>0$.

To compare sender values, observe that the maximal initial payoff ratio tends to $2-(r_1+\kappa)/2$. The lower bound follows by evaluating it at $b$; the upper bound follows from the uniform upper bound $s(\mu)\leq1-r_1/4+o(1)$ inside the band established in Appendix~\ref{app:concentration}. Hence
\[
V_1\to2\pi-\frac{r_1+\kappa}{2}\pi+\kappa\pi^2,
\qquad V_2\to2\pi^2.
\]
Their difference is \eqref{eq:sender-loss-margin}. Uniform strict margins prove all the claims for sufficiently small $\delta$. Their continuity in $r_2$ proves the final assertion of the proposition.

\section{The shared-verification continuation game}\label{app:sharing}
At an interior posterior and under an atomless $F$, $0<Q<1/2$, so $0<a=rQ<1$. If receiver $i$ assigns probability $h_i$ to no other receiver broadcasting, inspection is optimal exactly when $c_i\leq q(t_i,\mu)h_i$. Independence gives $p_i=ah_i$ and hence \eqref{eq:BNE}. The subtraction argument in the text rules out asymmetric equilibria. In the scalar equation \eqref{eq:p-scalar}, the left side is strictly increasing in $p$ and the right side is weakly decreasing. Their endpoint values give one solution. Cutoff indifferences have probability zero, establishing uniqueness up to null events. At $\mu\in\{0,1\}$ or $r=0$, nobody inspects and all expressions are interpreted by continuity.

Put $m=n-1$, $d=1+(n-2)\rho p$, and
\begin{equation}\label{eq:eta}
\eta=\frac{m\rho p}{d}\in[0,1).
\end{equation}
Differentiation of \eqref{eq:p-scalar} gives \eqref{eq:pderivatives} and
\begin{equation}\label{eq:hderivatives}
h_a=-\frac{\eta h}{a},\qquad
h_r=-\frac{\eta h}{r},\qquad h_\rho<0.
\end{equation}
The last strict inequality applies at interior posteriors and $r>0$, including the right derivative at $\rho=0$. For $\rho>0$, increasing $n$ also strictly lowers $p$ and $h$, by comparing the right sides of the scalar equations at the old solution.

The probability of remaining uninformed is $h(1-rqh)$, proving \eqref{eq:lambda-sharing}. Multiplying this by $q$ gives the decision loss; integrating costs up to $qh$ gives expenditure and \eqref{eq:sharing-welfare}. For fixed $c$, a receiver that does not inspect has expected loss $qh$, whereas one that does inspect has loss $c$, proving the expression $\min\{c,qh\}$. Because
\[
\frac{\partial L_t}{\partial h}=q(1-rqh)\geq0
\]
and equality for an interior type and posterior is possible only at $rq=h=1$, total loss is strictly increasing in $h$ on any nondegenerate interval. Wider sharing therefore strictly improves cost-averaged welfare. It weakly improves welfare at each fixed cost. Cheaper verification also improves welfare at a fixed public posterior:
\[
\frac{\partial L_t}{\partial r}
=q(1-rqh)h_r-\frac12q^2h^2<0.
\]
This fixed-posterior comparison does not incorporate the sender's disclosure response.

For completeness, cheaper verification raises each type's probability of final revelation at a fixed posterior. With $v_t=rqh$, differentiation yields
\[
(\lambda_t)_r=\frac hr\{\eta(1-v_t)+(1-\eta)v_t\}>0.
\]
In contrast, at a fixed $a$, averaging across thresholds gives $\lambda=1-h+ah^2$ and thus \eqref{eq:lambda-avg}. The comparison between information and sharing can change sign as $p$ falls. If $a>1/2$, the minimum revelation probability occurs when $p=1/2$, at
\[
\rho_*=2[1-(2a)^{-1/(n-1)}]\in(0,1).
\]
If $a\leq1/2$, aggregate final information weakly increases throughout. These are fixed-posterior comparisons.

Finally, aggregate acceptance conditional on final revelation is $\mu$. Averaging the unrevealed action and the revealed action gives
\[
s=F+(1-h)(\mu-F)+rh^2\int[\mu-\ind\{t\leq\mu\}]q\dd F,
\]
which is \eqref{eq:s-sharing}. In the smooth family this reduces to \eqref{eq:s-sharing-family}. Its derivative with respect to $h$ is $-\alpha zw(1-2rzh)$. Combining it with $h_\rho<0$ gives the pointwise sender-payoff comparison in the text.

\section{Global optimality with weak sharing and proof of Proposition~\ref{prop:opposite}}\label{app:opposite}
Fix $n$ and $\alpha>0$. Define
\[
W_\rho(\mu)=\frac{s(\mu;r,\rho,n)-(1-\alpha)\mu}{\alpha}-K\mu^2,
\qquad g_\rho(\mu)=\frac{W_\rho(\mu)}\mu,
\qquad \Psi_\rho=g_\rho'.
\]
The ratio is extended smoothly at zero. The equilibrium equation can be written as
\[
h=(1-\rho rQh)^{n-1}.
\]
The derivative of its left side minus right side with respect to $h$ is strictly positive. The implicit function theorem therefore gives a smooth solution in $\mu$ and the parameters near $\rho=0$, including at $Q=0$. Moreover, $W_\rho(0)=0$. The representation $g_\rho(\mu)=\int_0^1W_\rho'(t\mu)\dd t$ provides smooth extension and uniform derivative control at zero.

At $\rho=0$, $\Psi_0=T$, $\Psi_0'\leq-5/4$, and $\Psi_0(1)=-1-K<0$. On any fixed compact set of $(r,K)$, these strict inequalities persist for sufficiently small $\rho$, with $\Psi_\rho'<0$ everywhere and $\Psi_\rho(1)<0$. Therefore $g_\rho$ has at most one interior maximizer $y$. When it exists, the line $\mu g_\rho(y)$ strictly majorizes $W_\rho$ except at zero and $y$, and
\[
W_\rho''=2\Psi_\rho+\mu\Psi_\rho'<0\quad\text{for }\mu\geq y.
\]
If there is no positive root, the payoff is strictly concave. As in Appendix~\ref{app:smooth}, the unique optimum is $\Pi[0,y]$ when $\Psi_\rho(\pi)>0$, and pooling otherwise. This proves a global envelope characterization, not only persistence of a stationary candidate.

We now establish the strict signs in Proposition~\ref{prop:opposite}. At $\rho=0$,
\[
h_\rho=-(n-1)rQ,
\]
and differentiation of \eqref{eq:s-sharing} gives
\begin{equation}\label{eq:s-rho-zero}
s_\rho=(n-1)rQ(-D_F-2rH)
=(n-1)r\alpha z^2(1-2\mu)(1+\alpha z)(1-2rz).
\end{equation}
Consequently
\[
\left.\frac{\partial g_\rho(\mu)}{\partial\rho}\right|_{\rho=0}
=(n-1)rJ(\mu)(1+\alpha z)(1-2rz).
\]
Since $J'=\chi$ and $\chi(u)=0$, differentiating with respect to $\mu$ at $u$ yields \eqref{eq:sharing-contact}. All factors there are positive under \eqref{eq:opposite-r}. At $K=K_+$, the contact is $u>\pi$, and the implicit derivative satisfies
\[
y_\rho=-\frac{\partial_\rho\Psi_\rho}{\partial_y\Psi_\rho}>0.
\]
Choose $K<K_+$ sufficiently close to $K_+$. At $\rho=0$, the contact then lies strictly between $u$ and $1/2$, so $\chi(y)<0$ and $y_r<0$. The positive sharing derivative persists. The global-envelope argument above and continuity extend both strict signs to sufficiently small positive $\rho$.

At $\rho=0$, the envelope derivatives are
\[
V_r=\alpha\pi J(y)>0,
\qquad V_\rho=\alpha(n-1)r\pi J(y)(1+\alpha z(y))(1-2rz(y))>0.
\]
Here $y<1/2$ and $r$ satisfies the strict bound in the proposition. Both inequalities persist for sufficiently small positive sharing. The neighborhood of $K_+$ may depend on $n,\alpha,r,\pi$. After $K$ is fixed in that neighborhood, the admissible sharing bound $\bar\rho(K)$ may also depend on $K$. The argument does not give a common positive sharing bound as $K$ approaches $K_+$, or as the other primitives vary.

\section{Proof of Proposition~\ref{prop:sharing-loss}}\label{app:sharing-loss}
Part (i) follows from Appendix~\ref{app:sharing} and \eqref{eq:lambda-avg}. Conditional on a fixed posterior, the aggregate final law is $(1-\lambda)\delta_\mu+\lambda[(1-\mu)\delta_0+\mu\delta_1]$. It is strictly increasing in the Blackwell order with $\lambda$. Since $h_\rho<0$, its local direction is the opposite of the sign of $2p-1$.

For part (ii), we first establish a uniform sufficient condition for pooling. With two receivers, $h=(1+\rho rQ)^{-1}$. In the smooth family,
\[
|Q'|\leq1,\quad |Q''|\leq3,\quad |h'|\leq2,\quad |h''|\leq14,
\quad |(h^2)'|\leq4,\quad |(h^2)''|\leq36.
\]
The first bound follows from concavity of $Q$, symmetry, and $Q'(0)=1$; the second follows from $Q''=-2f_\alpha$ and $f_\alpha\leq3/2$. The bounds on $h$ follow by differentiating its explicit expression and using $\rho r\leq2$ and $h\leq1$.

Write $D_F=-\alpha z(1-2\mu)$ and $H=\alpha z^2(1-2\mu)$. Elementary differentiation gives the bounds
\[
|D_F|\leq\alpha/4,\quad |D_F'|\leq\alpha,\quad |D_F''|\leq6\alpha,
\quad |H|\leq\alpha/16,\quad |H'|\leq5\alpha/8,\quad |H''|\leq5\alpha.
\]
For example, $H'=\alpha(2z(1-2\mu)^2-2z^2)$ and $H''=\alpha[2(1-2\mu)^3-12z(1-2\mu)]$, from which the last two bounds follow. Hence
\begin{align*}
|(hD_F)''|&\leq14\alpha/4+4\alpha+6\alpha=27\alpha/2,\\
|r(h^2H)''|&\leq2(36\alpha/16+8\cdot5\alpha/8+5\alpha)=49\alpha/2.
\end{align*}
Equation \eqref{eq:s-sharing} yields $|s''|\leq38\alpha$, uniformly in $r,\rho,\mu$. Thus $\kappa>19\alpha$ makes $v=s-\kappa\mu^2$ strictly concave, and pooling is uniquely optimal for all the parameter values asserted.

At pooling, $a=a_\pi$ is unchanged when sharing changes. Formula \eqref{eq:lambda-golden} proves the exact information comparison. Inspection falls because $p$ falls, and the typewise welfare comparison follows from the strict fall in $h$. To verify nonemptiness analytically, choose
\[
\alpha>4\sqrt5-8,\qquad
\frac{8(\sqrt5-1)}{4+\alpha}<r<2.
\]
These intervals are nonempty for some $\alpha\leq1$. At $\pi=1/2$, the left side of \eqref{eq:golden} would equal $r(4+\alpha)/16$, strictly above its right side. Continuity therefore gives the desired strict inequality for all $\pi<1/2$ sufficiently close to $1/2$. This is an analytical existence argument and does not rely on selecting numerical parameter values.

\section{A large audience}\label{app:large}
The following result records the role of full-audience dissemination. It is separate from the local comparative statics in Proposition~\ref{prop:opposite}.

\begin{lemma}\label{lem:large}
Let $F$ be atomless with a twice continuously differentiable distribution function on $[0,1]$. Fix $\pi\in(0,1)$, $\kappa>0$, $r\in(0,2]$, and $\rho\in(0,1]$. The sender values the average action. For every sufficiently large audience size $n$, its unique optimal experiment is pooling. At that experiment, every threshold type's probability of learning the state tends to one, its expected loss tends to zero, and sender value tends to $\pi$. Writing $p_n$ for the equilibrium inspection probability,
\begin{equation}\label{eq:large-rates}
p_n\sim\frac{\log n}{\rho n},\qquad
np_n\sim\frac{\log n}{\rho},\qquad
L_n\sim\frac{\log n}{\rho r n},
\end{equation}
where $L_n$ is average receiver loss. Total inspection expenditure tends to zero.
\end{lemma}

\begin{proof}
At every interior posterior, $a=rQ>0$. Equation \eqref{eq:p-scalar} implies that $p_n$ and $h_n=p_n/a$ decrease with $n$. Their limits must be zero: a positive limiting $p_n$ would force $(1-\rho p_n)^{n-1}$ to zero. The continuous functions $h_n$ therefore converge uniformly to zero on any compact interior posterior interval by monotone convergence of continuous functions. Let
\[
e_n(\mu)=h_nD_F(\mu)+rh_n^2H(\mu),
\qquad s_n(\mu)=\mu+e_n(\mu).
\]
Since $D_F$ and $H$ vanish at both endpoints and $0\leq h_n\leq1$, interior convergence also implies $\|e_n\|_\infty\to0$ on $[0,1]$.

Uniform value convergence alone would not establish eventual exact pooling. We also need local curvature control. Regarding $h_n$ as a function of $a$, \eqref{eq:eta} gives
\[
(h_n)_a=-\frac{\eta h_n}a,\qquad
a\eta_a=\frac{\eta(1-\rho p_n)}{d^2}\leq\eta.
\]
A second derivative yields
\[
(h_n)_{aa}=\frac{h_n}{a^2}(\eta^2+\eta-a\eta_a),\qquad
0\leq(h_n)_{aa}\leq\frac{2h_n}{a^2}.
\]
On a neighborhood of $\pi$, $a$ is bounded away from zero and its first two posterior derivatives are bounded. It follows that $h_n$, $h_n'$, and $h_n''$ converge uniformly to zero there. The regularity of $F$ gives the same local convergence for $e_n,e_n',e_n''$.

The gap between the tangent to $v_n(\mu)=\mu-\kappa\mu^2+e_n(\mu)$ at $\pi$ and $v_n(\mu)$ itself is
\begin{equation}\label{eq:large-tangent}
\kappa(\mu-\pi)^2+e_n(\pi)+e_n'(\pi)(\mu-\pi)-e_n(\mu).
\end{equation}
Choose a fixed neighborhood $|\mu-\pi|<\varepsilon$ within $(0,1)$. For large $n$, $\sup|e_n''|<\kappa$ there; Taylor's theorem bounds \eqref{eq:large-tangent} below by $\kappa(\mu-\pi)^2/2$. Outside that neighborhood it is bounded below by
\[
\kappa\varepsilon^2-2\|e_n\|_\infty-|e_n'(\pi)|>0
\]
for large $n$. Thus the tangent is a strict global majorant away from $\pi$, proving unique pooling.

At pooling, set $a=rQ(\pi)>0$ and $m=n-1$. Since $h_n\to0$, \eqref{eq:lambda-sharing} and \eqref{eq:sharing-welfare} give the typewise limits. Also $s_n(\pi)\to\pi$, and the public information cost is zero.

For the rate calculation, let $z_n=m\rho p_n$. From $1-x\leq e^{-x}$ and the equilibrium equation,
\[
z_ne^{z_n}\leq m\rho a,
\]
so $z_n=O(\log n)$ and $np_n^2\to0$. Moreover, $np_n\to\infty$; otherwise $h_n=(1-\rho p_n)^{n-1}$ would stay bounded away from zero along a subsequence. Taking logarithms and using $m p_n^2\to0$ gives
\[
\log z_n+z_n=\log(m\rho a)+o(1).
\]
Therefore $z_n\sim\log n$ and the first two rates in \eqref{eq:large-rates} follow. If $Q_2=\int q(t,\pi)^2\dd F(t)$, then
\[
L_n=Qh_n-\frac r2Q_2h_n^2
=\frac{p_n}{r}-\frac{Q_2p_n^2}{2rQ^2}.
\]
This proves the loss rate. Total inspection expenditure is $nQ_2p_n^2/(2rQ^2)$, which tends to zero. The result concerns loss per receiver; it does not assert that the sum of decision losses converges to zero.
\end{proof}

\end{document}